\documentclass[twocolumn,amsmath,amssymb,groupedaddress,superscriptaddress]{revtex4}
\usepackage{graphicx}
\usepackage{bm}
\usepackage{dsfont}

\newcommand{\ket}[1]{\mbox{$| #1 \rangle$}}

\usepackage{mathptmx} %comapct fonts

\begin{document}
\preprint{}
\title{Quantum Metrology in Non-Markovian Environments}
\author{Alex W. Chin}
\address{Institute f{\"u}r Theoretische Physik und Center for Integrated Quantum Science and Technology IQST, Universit{\"a}t Ulm,
89069 Ulm, Germany}
\address{Theory of Condensed Matter Group, University of Cambridge,
J J Thomson Avenue, Cambridge, CB3 0HE, United Kingdom}
\author{Susana F. Huelga}
\address{Institute f{\"u}r Theoretische Physik und Center for Integrated Quantum Science and Technology IQST, Universit{\"a}t Ulm,
89069 Ulm, Germany}
\author{Martin B. Plenio}
\address{Institute f{\"u}r Theoretische Physik und Center for Integrated Quantum Science and Technology IQST, Universit{\"a}t Ulm,
89069 Ulm, Germany}

\date{\today}

\begin{abstract}
We analyze precision bounds for local phase estimation in the presence of general, non-Markovian phase noise. We demonstrate that the metrological equivalence of
product and maximally entangled states that holds under strictly Markovian dephasing fails
in the non-Markovian case. Using an exactly solvable model of a
physically realistic finite band-width dephasing environment, we demonstrate that
the ensuing non-Markovian dynamics enables quantum correlated states to outperform
metrological strategies based on uncorrelated states using otherwise identical resources. We show that this conclusion is a direct result of the coherent dynamics
of the global state of the system and environment and therefore the obtained scaling with the number of particles, which surpasses the standard quantum limit but does not achieve Heisenberg resolution, possesses
general validity that goes beyond specific models. This is in marked contrast with the situation encountered under general Markovian noise, where an arbitrarily small amount of noise is enough to restore
the scaling dictated by the standard quantum limit.
\end{abstract}


\maketitle
Entangled states can achieve a resolution in metrology
that surpasses
the precision limits achievable with uncorrelated probes, a significant
result of both fundamental and practical relevance first put forward by
Caves \cite{caves}. The potential usefulness of entangled states in
overcoming the shot noise limit in precision spectroscopy (also referred to as standard quantum limit or standard scaling) was proposed
in \cite{nist}, and the first experimental results concerning precision
measurements using entangled input states have been presented recently
\cite{blatt}. However, the saturation of the Heisenberg limit by maximally
entangled states assumes a fully coherent evolution, whereas in real
experiments there will always be some degree of decoherence or a limitation
on the total time over which measurements can be performed. Precision
spectroscopy in the presence of Markovian dephasing was first analyzed
in \cite{us}, where it was shown that given a fixed number of particles
$n$ and a total available time $T$ for the frequency estimate to be
completed, uncorrelated and maximally entangled preparations of $n$
particles achieve exactly the same precision when subject to Markovian
dephasing. Hence these two preparations are {\em metrologically} equivalent
in those circumstances.

Here, we analyse if this equivalence persists when the system is subject to non-Markovian noise. Under the same rules as above, namely fixed $n$ and $T$, we show that
in the presence of realistic, finite temperature, finite bandwidth environments,
a measurement strategy can always be found in which the use of $n$-particle
entangled states leads to a lower frequency uncertainty when compared
to the use of $n$ uncorrelated input states. Moreover, we demonstrate on very
general grounds that for these strategies the ratio between the optimal
resolution of entangled and uncorrelated probes obeys a characteristic
power law $\propto n^{1/4}$. These results imply that entangled states
{\em can} be used to gain an advantage for precision measurements in
the presence of noise, and that entanglement-enhanced metrology could
be practically implemented in wide variety of condensed matter systems
such as realizations of solid-state qubits and biomolecular systems which
are typically subject to non-Markovian environments characterized by long
correlation times and/or structured spectral features \cite{nv}.

%%%%%%%%%%%%%%%%%%%%%%%%% part 2 %%%%%%%%%%%%%%%%%%%%%%%
To show this, let us consider a system Hamiltonian $\omega_0\sigma_z$ which
is subject to a system-environment interaction that induces pure dephasing,
the form of noise that tends to manifest at the shortest time scales in most
qubit realizations. In this case, the coupling to the environment is of the
form $\sigma_z\otimes B$, where $B$ is some operator only including bath degrees
of freedom. Then, denoting by $(\ket{1},\ket{0})$ the eigenbasis of $\sigma_z$, quite generically, the time evolution of the reduced density
matrix of the system satisfies
\begin{eqnarray}
    \rho_{ii}(t) &=& \rho_{ii}(0) \;\; \mbox{\rm{for}} \;\; (i=0,1)\\
    \rho_{01}(t) &=& \rho_{01}(0)e^{-2\gamma(t)}.
\end{eqnarray}
Now we consider a typical Ramsey spectroscopy set-up for $n$ uncorrelated
particles \cite{wineland} to find that the resulting single particle signal
is given by
\begin{equation}
    p_0=\frac{1}{2} \left(1+\cos(\phi t) e^{-\gamma (t)}\right),
\end{equation}
where $\phi$ is the detuning between the frequency $\omega$ of the external oscillator and the atomic frequency $\omega_0$ to which we intend to lock it to and $t$ is the time between Ramsey pulses \cite{us}. Using the same notation as
in \cite{us}, the best resolution in the estimation of $\omega_0$ is given
by the expression
\begin{equation}
    \delta \omega_0^2=\frac{1}{N F(\phi)},
    \label{signal}
\end{equation}
where $N$ is the total number of experimental data ($N=(T/t)\,n$) and $F$ is the so-called Fisher information \cite{bcm}. This quantity can be easily evaluated in our case as
\begin{equation}
F(\phi) = \sum_{i=0,1} \frac{1}{p_i} \left(\frac{\partial p_i}{\partial \phi}\right)^2.
\end{equation}
We then find the frequency uncertainty to be
\begin{eqnarray}
    \delta\omega_0^2 = \frac{1-\cos^2(\phi t)e^{-2\gamma (t)}}{nTt\sin^2(\phi t)e^{-2\gamma (t)}}\label{domega}.
\end{eqnarray}
We wish to determine the best operating point $\phi$ and the best
interrogation time $t_{u}$ which minimize Eq. (\ref{domega}), as these
two quantities are under experimental control. To this end, one computes
the derivatives of  $\delta\omega_0^2$ with respect to $\phi$ and $t$
and then equates these derivatives with $0$. Independently of the choice of
$\gamma(t)$, we conclude from the derivative of $\delta \omega_0^2$ with
respect to $\phi$ that
\begin{eqnarray}
    \phi \, t_{u} = \frac{k\pi}{2}\label{Constraint1}
\end{eqnarray}
for odd $k$ or, in other words, the choice that ensures $\cos{\Delta t_{u}}=0$
is optimal. Inserting $\phi t_{u} = \frac{k\pi}{2}$ in the expression for
the derivative with respect to $t_{u}$ to eliminate $\phi$, these expressions
simplify considerably and we obtain the second constraint
\begin{equation}
    2t\frac{d\gamma(t)}{dt}|_{t=t_{u}} = 1.\label{ligo1}
\end{equation}
Using the Eq. (\ref{Constraint1}) in Eq. (\ref{domega}) we have
\begin{equation}
    \delta\omega_0^2|_{u} = \frac{1}{nTt_{u}}e^{2\gamma(t_{u})},\label{optomega}
\end{equation}
where the optimal interrogation time $t_{u}$ is determined by Eq. (\ref{ligo1}). The Markovian case is recovered from these equations for $\gamma(t)=\gamma(0)\,t$
and the expressions above reduce to those presented in \cite{us}.
% To solve these equations for a given $\gamma(t)$ one first solves Eq.
% (\ref{Constraint2}) to determine the optimal $t_{u}$ and then uses Eq.
% (\ref{Constraint1}) to determine the value of $\Delta$.

An analogous calculation can be done for an initial preparation of $n$ particles
in a maximally entangled state $|0\rangle^{\otimes n}+|1\rangle^{\otimes n}$,
leading to the result that the optimal frequency resolution is
\begin{equation}
    \delta\omega_0^2|_{e} = \frac{1}{n^2Tt_{e}}e^{2n\gamma(t_{e})},\label{optdeltaent}
\end{equation}
where the optimal interrogation time for {\em entangled} particles $t_{e}$ is determined by the constraint,
\begin{equation}
    2n t\frac{d\gamma(t)}{dt}|_{t=t_e} = 1.\label{ligo2}
\end{equation}
In the Markovian case, the additional factor of $n$ in the denominator of Eq. (\ref{optdeltaent}) is canceled out due to an accompanying decrease in $t_{e}$ by a factor of $n$ relative to $t_{u}$. The optimal frequency resolution is therefore identical to that obtained with uncorrelated particles, and thus maximally entangled and uncorrelated states are {\em metrologically} equivalent in the presence of local Markovian dephasing. Although Markovian dephasing does not allow any advantage to be gained from using maximally entangled states, the conclusions drawn above are very general, as the expressions above do not depend on the precise form of the decoherence model. Provided that it generates Markovian dephasing, the bath operator $B$ could be highly non-linear, with a complex spectral structure, quantum or classical.

We now move beyond the standard Markovian treatment and consider the performance of maximally entangled states in the presence of non-Markovian dephasing. We shall first study some specific, exactly solvable models, which demonstrate that entangled and uncorrelated probes are no longer metrologically equivalent in the presence of non-Markovian dynamics, and then discuss why this result is in fact independent of the microscopic details of the environment for most realistic system-bath structures.

{\em An exactly solvable model --}
Let us first consider the exactly solvable model (independent boson model)
\cite{breuer}. Here the bath operator $B$ is simply a sum of linear couplings
to the coordinates of a continuum of  harmonic oscillators described by a
spectral function $J(\omega)$ \cite{breuer,leggett,weiss}. Then we have
\begin{eqnarray}
    \gamma(t) = \frac{1}{2}\int_0^{\infty} d\omega J(\omega) \coth\left(\frac{\omega\beta}{2}\right)
    \frac{1-\cos(\omega t)}{\omega^2}.\label{gammaint}
\end{eqnarray}
where $\beta$ is the inverse temperature.\\
{\em Power-law spectral densities with exponential cut-offs --}
The coupling to a bath of harmonic oscillators is the most common setting used
in the study of open-quantum systems, and an extremely large number of physical environments can be described by a general power-law form for the spectral density \cite{leggett,weiss}. Following Ref. \cite{breuer}, we therefore consider $J(\omega)=\alpha \omega_c^{1-s}\omega^s e^{-\omega/\omega_c}$, where $\alpha$ is a dimensionless coupling constant and $\omega_{c}$ cuts off the spectral density at high frequencies. For zero temperature, $t>0$ and $s>0$, we obtain the result
\begin{eqnarray}
    \gamma(t) = \frac{\alpha}{2}\left(1-\frac
    {\cos[(s-1) \tan^{-1}(\omega_c t)]\, \Gamma(s-1)}{(1+\omega_c^2 t^2)^{\frac{s-1}{2}}}\right),\nonumber\\ \label{gamma}
\end{eqnarray}
where $\Gamma(s-1)$ is the Euler Gamma function. Taking the limit $s\rightarrow1$ carefully, one also finds
\begin{eqnarray}
    \gamma(t,s=1) = \frac{\alpha}{2}\ln(1+\omega_c^2 t^2).\label{ohmicgamma}
\end{eqnarray}
From Eq. (\ref{ohmicgamma}) one immediately sees that at short ($\omega_{c}t\ll1$) and long ($\omega_{c}t\gg1$) times,  $\gamma(t)$  has a power law dependence on time, and it is therefore instructive to analyse a generic $\gamma(t)$ of the from $\gamma(t)=\alpha t^\nu$.  We define the relative frequency resolution of entangled and uncorrelated probes $r=|\delta \omega_0|_{u}/|\delta \omega_0|_{e}$. We then find
\begin{equation}
r^2=n\left(\frac{t_{e}}{t_{u}}\right) e^{2\gamma(t_{u})-2n\gamma(t_{e})}.\label{r}
\end{equation}
In the absence of dephasing noise, $r=\sqrt{n}$ (Heisenberg limit), while in
the Markovian case the metrological equivalence of the correlated and entangled
probes is presented by the result $r=1$. Using the constraint equations
Eq. (\ref{ligo1}) and (\ref{ligo2}), it can be seen that for
the general power law form of $\gamma(t)=\alpha t^{\nu}$, we always obtain $\gamma(t_{u})=n\gamma(t_{e})$ and the exponential term in Eq. (\ref{r})
always equals unity. Hence $r$ is determined by the ratio of best interrogation times
$t_{u}/t_{e}$. Similarly,
one can show that the ratio $t_{u}/t_{e}=n^{1/\nu} $ and therefore
$r^2=n^{\frac{\nu-1}{\nu}}$. From this result we see that only for $\nu>1$
there is an advantage in using entangled probes, and $r$ approaches the Heisenberg
limit from below as $\nu\rightarrow\infty$. The case of $\nu=1$ corresponds to
the Markovian case, whilst $\nu<1$ always favours uncorrelated probes.

With this analysis we can use Eq. (\ref{gamma}) to assess $r$ as a function of the bath exponents $s$. For short times, expanding Eq. (\ref{gamma}) to the leading-order in $\omega_{c}t$, it can be seen that for all spectral densities $\gamma(t)\propto t^2$, and one then obtains $r=n^{\frac{1}{4}}$.  The necessary interrogation times for entangled states satisfy $t_{e}\propto (\omega_{c} \sqrt{n})^{-1}$, which is consistent with the short time approximation of $\gamma(t)$.
%
% While such fast measurements might be difficult to achieve in certain systems,
% a natural advantage of short interrogations is that the measured signals are
% strong due to the short time over which the signal decoheres.
In many cases, and particularly in molecular and magnetic systems, the conditions
on the measurement time may be met easily with current experimental methods due
to the sluggishness of the dephasing environments. We also note that in the limit
of a static bath which induces Gaussian inhomogeneous broadening,
$\gamma(t)\propto t^2$ even for long times \cite{inhomogeneous}.

For times much greater than $\omega_{c}^{-1}$, we find that
$\gamma(t)\propto t^{1-s}$ for $0<s<1$.  For this case, known as
sub-Ohmic dissipation \cite{leggett,weiss}, uncorrelated probes are
always favoured, while for $s=1$ we can analytically evaluate the optimal interrogation times for each initial preparation without considering the
long or short time limits. The exact result is
\begin{equation}
    r =\sqrt{n} f(\alpha,n),
\end{equation}
where
\begin{equation}
    f(\alpha,n)=\sqrt{\left[\frac{(2 \alpha/(2 \alpha-1))^{\alpha}}{(2 n \alpha/(2 n \alpha-1))^{n \alpha}}\right] \sqrt{(2 \alpha-1)/(2 n \alpha -1)}},
\end{equation}
and $\alpha>1/2$ \cite{foot}. The results are shown in Figure 1, illustrating that maximally entangled states in the presence of zero temperature Ohmic baths outperform uncorrelated probes for any $n$, with  $r\rightarrow n^{\frac{1}{4}}$ as $n\rightarrow\infty$ and/or $\alpha\rightarrow\infty$.\\

\begin{figure}
\begin{center}
\includegraphics[width=0.40\textwidth]{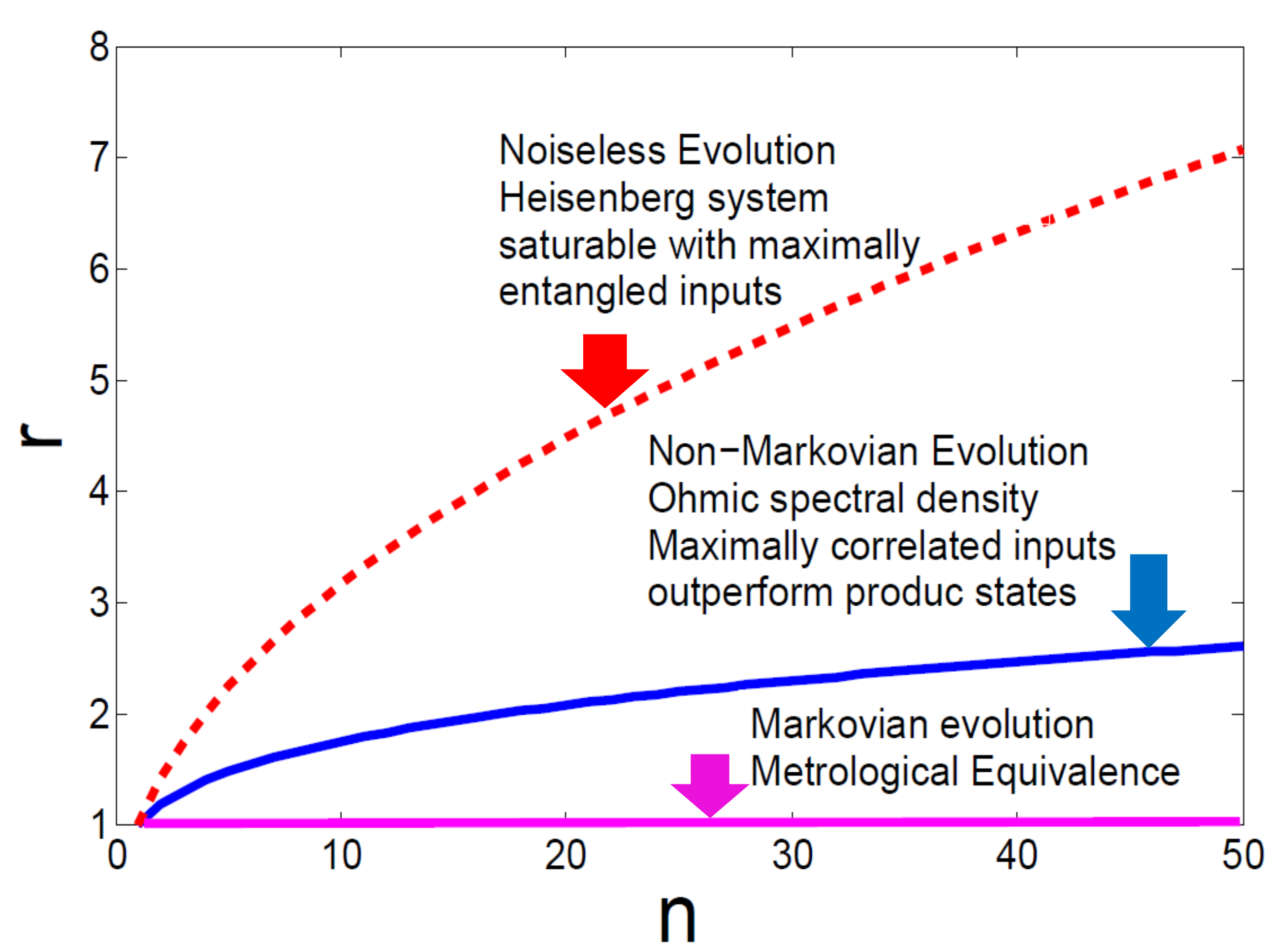}\vspace{-0.7cm}
\end{center}
\caption{
Ratio $r$ between the optimal resolution achievable with uncorrelated and maximally entangled inputs as a function of the number of particles $n$. The dashed line shows the expected behaviour in the absence of
noise where $r=\sqrt{n}$ (Heisenberg limit), while $r$ becomes equal to 1 (pink line) when the noise is fully Markovian. In the presence of non-Markovian phase decoherence, product states and maximally entangled initial preparations are no longer metrologically equivalent.
In the case of a zero temperature bath with an Ohmic spectral density ($s=1$),
maximally entangled states allow for a higher resolution for any value of $n$ and $r$ displays a typical $n^{1/4}$ dependence as shown by the solid line in the figure.}
\end{figure}

{\em Lorentzian spectral density --} Now we consider the spectral
density
\begin{displaymath}
    J(\omega) = \frac{1}{\pi} \frac{ag}{g^2 + \omega^2},
\end{displaymath}
where $a$ regulates the coupling strength. We then find for $T=0$ that
\begin{eqnarray}
    \gamma(t) = \frac{a}{4g}(\frac{1}{g}(e^{-gt}-1) + t)
\end{eqnarray}
for $g\ge 0$ and $t\ge 0$. Now inserting the necessary condition
$\Delta t = \pi/2$ in the expression for $\delta\omega_0^2$ we obtain
\begin{displaymath}
    \delta\omega_0^2|_{u} = \frac{1}{nTt}e^{\frac{a(-1+e^{-gt}+gt)}{2g^2}}.
\end{displaymath}
The second necessary condition for an optimum imposes that the optimal
time satisfies
\begin{displaymath}
    at(1-e^{-gt}) = 2g.
\end{displaymath}
This is a transcendental equation but, if we are interesting in the short
time behaviour $gt\ll 1$, then we find in lowest order as an approximate solution
\begin{displaymath}
    t^2 = \frac{2}{a}
\end{displaymath}
and employing the Newton method on the function $f(t)=at(1-e^{-gt}) -2g$
with starting point $t = \sqrt{\frac{2}{a}}$ we find the improved value
\begin{displaymath}
    t = \sqrt{\frac{2}{a}}\left(1+\sqrt{\frac{g^2}{8a}}\right).
\end{displaymath}
Inserting this into the expression for $\delta\omega_0^2$ we find
\begin{displaymath}
    \delta\omega_{opt}^2|_{u} = \frac{1}{nT}\sqrt{\frac{a}{2}}\frac{\sqrt{8a}}{\sqrt{8a}+g} e^{\frac{g}{3}\sqrt{\frac{2}{a}}-1}
\end{displaymath}
Repeating the calculation for a maximally entangled state, we obtain in the short time limit $gt\ll1$,
\begin{displaymath}
    \delta\omega_{opt}^2|_{e} = \frac{1}{nT}\sqrt{\frac{a}{2n}}\frac{\sqrt{8an}}{\sqrt{8an}+g} e^{\frac{g}{3}\sqrt{\frac{2}{an}}-1}.
\end{displaymath}

We find an improved precision for maximally entangled states as
$\delta\omega_{opt}^2$ is reduced by a factor $\sqrt{n}$ whenever
$8an\gg g^2$. If that last condition is not satisfied, the above
approximate expressions fail to hold, as then $g$ becomes large.
A numerical calculation reveals that for $8an\ll g^2$, maximally
entangled and product states achieve the same precision and the
optimal interrogation time becomes large. That entangled and product
states then achieve the same precision can be expected as memory
effects in the bath become negligible for large interrogation
times.

{\em Beyond specific models --}
The key point illustrated by the examples above is that maximally entangled
states achieve their optimal interrogation time at shorter time intervals
than uncorrelated states and can hence benefit more from non-Markovian
noise features. This is due to the characteristic behaviour $\gamma(t)\propto t^2$
which governs short times in the models above, and which leads to a decrease
in the optimal interrogation time for entangled particles that only scales
as $n^{-\frac{1}{2}}$ (c.f. $t_{e}\propto n^{-1}$ for the Markovian case).

However, the quadratic behaviour
of $\gamma(t)$ is not a specific feature of our chosen noise model, but rather
a general consequence of the unitary evolution of the total system and environment
state.  The essential observation is that the function $\gamma(t)$ appears in
the dynamics of the sub-system as the result of transitions induced in the bath
by the system-bath interaction. At a short time $t$ after the system-bath
interaction is switched on, the probability for the bath state to make a transition
to any state orthogonal to its initial condition is {\em always} proportional
to $t^2$.  This universal time dependence for quantum mechanical transitions is
the fundamental basis of the quantum Zeno effect, and has been extensively and rigourously investigated \cite{peres,rimini}. Hence, for essentially {\em all} noise sources treated within the standard framework of open-quantum system theory, entangled-state input protocols can always be found which outperform uncorrelated probes, whatever the microscopic details of the bath and the system-bath interaction.

This general result leads to the concept of a new fundamental limit for quantum metrology in the presence of noise, which for simplicity we shall refer as the
{\em Zeno limit}. For sufficiently fast interrogation times, we find the model-independent scaling law  for the Zeno limit $r=n^{\frac{1}{4}}$, which
is below the Heisenberg limit $r=\sqrt{n}$, but always above the Markov limit
$r=1$.  For the specific noise models studied above, we also find that $t_{e}$
can be simply related to $r$ through the relation $r^2\omega_{c}t_{e}=1$ at
$T=0$ K. Again, given the universal scaling law for $r$, a relation of this
form should also be expected to hold for other noise models, except that
$\omega_{c}$ should be replaced by the fastest dynamical frequency of the
environment in these models. It is worth noting that if the effect of decoherence if formally thought of as
the action of environmental projective measurements, our result showing a ratio $t_e/t_u=1/\sqrt{n}$ for the optimal interrogation time of maximally entangled and product states is in agreement with recent work
deriving the time scale for quantum Zeno dynamics in terms of the Fisher information \cite{smerzi}.\\

{\em Finite Temperatures --} The arguments given above also naturally apply to
the case of finite temperatures,  where again we find that a Zeno-limit emerges. However, the typical energy scale that determines the
optimal interrogation time $t_{e}$ now depends explicitly on temperature.
This can be seen directly in the high temperature limit of our exact model,
where the factor $\coth(\beta\omega_{c}/2)$ in Eq. (\ref{gamma}) can be expanded
to leading order in $\beta\omega_{c}$ over the entire integration range. For an
Ohmic bath this leads to $\gamma(t) = \alpha\beta^{-1}(t\tan^{-1}(\omega_{c}t)-\ln(\sqrt{1+\omega_{c}^{2}t^2}) \,\omega_{c}^{-1})$. Again, a Zeno-limit appears at short times with $\gamma(t)\approx\alpha\beta^{-1}\omega_{c}t^2/2$, which leads to the result $t_{e}=\sqrt{\frac{\beta}{4\alpha n \omega_{c}}}$. This result, derived in the high temperature limit, is consistent with our notion that it is the fastest time scale of the bath dynamics that sets the scale for the Zeno-limit interrogation time.
If the system is interrogated slower than this timescale, we find that entangled
and uncorrelated probes become equivalent again as $\gamma(t)\approx \alpha\beta^{-1}\omega_{c}t$ at long times and the Markov result is recovered.\\
One question remains though and that concerns the evaluation of the optimal resolution achievable in the presence of a given form of non-Markovian dephasing
provided that both the initial state preparation and the final measurement can be optimized. This is likely to be a complicated question, and in fact it has taken
almost 15 years to rigourously prove that, in the purely Markovian case, as argued in \cite{us}, the improvement obtained by using partially entangled states over product states and projective measurements is a mere
constant equal to $\sqrt{e}$ \cite{brazil}. We believe that a similar situation will be encountered in the present case, so that the scaling $n^{1/4}$ will be robust and
optimized preparations and measurements will determine the exact value of the multiplicative factor to be of the order of 1. We leave this as an open question and suggest that the use of convex
optimization techniques \cite{rafal} may help to prove this conjecture for those noise models whose effect can be represented as a completely positive and trace preserving (CPT) quantum channel.

{\em Conclusions -- } Using an exactly solvable model of non-Markovian dephasing,
we have shown that entangled probes can outperform uncorrelated preparations provided
the system is interrogated on time scales faster than the characteristic
frequencies of the bath dynamics. This conclusion holds for both zero and
finite temperatures, and is also valid for any other noise model arising from
an open-quantum system structure. This result can be naturally understood as
emerging from the scaling $t_{e}\propto n^{-1/2}$ in the number of correlated
particles, which causes the entangled probes to experience a suppressed level
of decoherence relative to the uncorrelated case, which in turn have to be measured at
much longer times. Thus we argue that the result $r=n^{1/4}$ for rapid measurements
is a new, fundamental metrological limit for entangled particles subject to independent
non-Markovian decoherence sources. We should stress that this result is in sharp contrast with the situation encountered
in the presence of general Markovian noise, where an arbitrarily small noise level is enough to restore the standard scaling \cite{rafal}.
Beyond the theoretical interest, we should stress the immediate practical relevance of our analysis, as the properties of non-Markovian noise which are crucial for obtaining the $n^{1/4}$ scaling are extremely generic and will be found in almost any realistic open quantum system. This work shows that an advantage can be obtained in real-world systems with a relatively simple, intuitive preparation and measuring protocol, and considerably expands the number of systems in which quantum metrology could be pursued.
Moreover, at the heart of this theory is the notion of probing the system on times which are faster than the typical memory times (assumed infinitely fast in the Markovian case) of the environment. In this regime, which we refer to as the Zeno limit, the metrological scaling advantage appears due to the characteristic time-dependence of coherently-evolving transition probabilities, which develop like $t^2$. This is a consequence of the standard microscopic model of open quantum systems, which posits that the total state of the system and environment evolves coherently, and that decoherence only emerges after the bath is traced over on time scales longer than the memory time. From the point of view of open quantum system theory, observing the $n^{1/4}$ scaling in metrology verifies the microscopic picture of how decoherence and dissipation emerge in small quantum systems.

{\em Acknowledgements -- } We are grateful to Fedor Jelezko and Rafal Demkowicz-Dobrzanski for comments and
to Simon Benjamin for drawing our attention to recent independent
results \cite{benja} where the same scaling emerges from somewhat different
models thus providing further evidence in favour of our conjecture of  a
novel fundamental metrological limit. Financial support
from the EU Integrated Project
Q-Essence, STREP action PICC, and an Alexander von Humboldt Professorship is gratefully acknowledged. AWC
acknowledges support from the Winton Programme for
the Physics of Sustainability.


\begin{thebibliography}{00}
\bibitem{foot} Note that $t_{c}$ does not exist below this coupling.
\end{thebibliography}
\end{document}